# Predicting Patient Recovery or Mortality Using Deep Neural Decision Tree and Forest

Mohammad Dehghani[1,*], Mohadeseh Zarei Ghobadi[2], Mobin Mohammadi[3], Diyana Tehrany Dehkordy[4]

[1] School of Electrical and Computer Engineering, University of Tehran, Tehran, Iran

[2] Department of Virology, School of Public Health, Tehran University of Medical Sciences, Tehran, Iran

[3] Isfahan University of Medical Sciences, Isfahan, Iran

[4] Department of Medical Informatics, Mashhad University of Medical Sciences, Mashhad, Iran

*Correspondence: dehghani.mohammad@ut.ac.ir

## Abstract

**Objective:** Identifying patients at high risk of mortality is crucial for emergency physicians to allocate hospital resources effectively, particularly in regions with limited medical services. This need becomes even more pressing during global health crises that lead to significant morbidity and mortality. This study aimed to present the usability deep neural decision forest and deep neural decision tree to predict mortality among Coronavirus disease 2019 (COVID-19) patients. To this end, We used patient data encompassing Coronavirus disease 2019 diagnosis, demographics, health indicators, and occupational risk factors to analyze disease severity and outcomes. The dataset was partitioned using a stratified sampling method, ensuring that 80% was allocated for training and 20% for testing. Nine machine learning and deep learning methods were employed to build predictive models. The models were evaluated across all stages to determine their effectiveness in predicting patient outcomes.

**Results:** Among the models, the deep neural decision forest consistently outperformed others. Results indicated that using only clinical data yielded an accuracy of 80% by deep neural decision forest, demonstrating it as a reliable predictor of patient mortality.

Moreover, the results suggest that clinical data alone may be the most accurate diagnostic tool for predicting mortality.

**Keywords:** Machine Learning, Deep Learning, Deep Neural Decision Forest, Mortality prediction, Healthcare.

## Introduction

Healthcare system inequality has long been a critical challenge, particularly in regions with low socioeconomic status where access to quality healthcare resources is limited (1). The emergence of health crises, such as pandemics or sudden surges in disease prevalence, has amplified concerns about the strain on healthcare systems. In these circumstances, there is an urgent need to optimize resource allocation (2), as the potential for shortages of healthcare professionals, hospital capacity, and medical supplies becomes more pronounced (3, 4). Predicting patient outcomes, such as mortality or recovery, plays a central role in enabling public health experts and hospital administrators to allocate resources more effectively (5).

Traditional expert scoring systems, which are frequently used in clinical settings to assess in-hospital mortality, provide valuable insights but also present limitations (6). These systems often rely on a fixed set of variables, which reduce their accuracy and adaptability in complex or rapidly evolving medical situations (7). This indicates the need for systems that can help prioritize patients and predict the outcomes of various treatments.

Artificial intelligence has emerged as a promising tool to address these challenges across a range of diseases to predict outcomes, support diagnosis, and optimize treatment strategies (8-11). In the context of infectious diseases, AI approaches have been used to predict mortality, identify risk factors, and assist in clinical decision-making (12).

The effectiveness of computed tomography (CT) scan and Reverse transcription polymerase chain reaction (RT-PCR) for Coronavirus disease 2019 (COVID-19) detection has been compared in several studies. Waller et al. reviewed diagnostic methods and found that CT scans are less sensitive and specific for COVID-19 compared to RT-PCR (13). They recommend using chest CT as a supplementary tool, particularly for patients who shows symptoms. Al-Momani proposed that RT-PCR remains the recommended first-line diagnostic method for COVID-19, whereas chest CT is not reliable enough for primary screening (14). Based on these studies, RT-PCR should be conducted in all cases with suspicious symptoms. However, it is important to note that a negative RT-PCR result does not always rule out COVID-19, and repeated testing may be necessary.

For mortality prediction of COVID-19, Queipo, et al explored seventy-one features using machine learning techniques, the variables that had the greatest influence on predicting mortality in this population were lymphocyte count, urea levels, $FiO_2$, potassium, and serum pH (15). The XGBoost method demonstrated the highest accuracy, achieving a precision of over 95%. A study was conducted by Talkhi et al. on 26,867 hospitalized individuals who tested PCR positive for COVID-19 (16). The results showed that the artificial neural network model had a slightly higher accuracy (90.27%) for predicting mortality compared to the linear regression model (90.15%). Moreover, individuals with chronic diseases such as cancer, kidney and blood diseases, and immunodeficiency were found to be at a higher risk of mortality.

Most of these studies focus on patients with specific characteristics (17) or use of CT imaging features (18) or tests, thereby limiting their applicability in certain circumstances. Moreover, there is no comparison between the importance of collected features and the impact of their existence or absence in prediction. Most deep learning algorithms proposed in studies lack interpretability, which is a crucial aspect of medical applications. To address these gaps, this study proposes an approach that combines the strengths of both machine learning and deep learning. By integrating an interpretable machine learning method with a high-performing deep learning technique, the research aims to develop a model that offers both accurate predictions and clear insights into the factors driving those predictions. In the following, we explained data, the preprocessing steps, and feature selection in the Materials and Methods.

Further, we evaluated the proposed models in the Results. Next, we discuss the outcomes in Discussion.

## Materials and Methods

Figure 1 illustrates the proposed model, which utilizes a COVID-19 patient dataset provided in our pervioue study (1). During preprocessing, we applied Min-Max normalization to the age column to scale the data to a range of 0 to 1. Categorical features were converted to numerical data using one-hot encoding. Missing values were imputed with appropriate values. Feature selection was then performed which involved calculating the frequency of each sign and symptom, with a threshold set at 0.1 to identify relevant features. Because the more repetitions of a feature, the more effective it is. The final selected features included Test-Result, Conformation-Method, Age, Ventilator, Cough, Apnea, Carcinoma, Healthcare-staff, and Intensive care units (ICU)-hospitalization (Table 1).

Table 1. The final selected features.

| | Selected Feature | Definition |
|---|---|---|
| 1 | Test-Result | Can be Positive or Negative.<br>Positive indicates that based on Clinical or RT-PCR, the patient has COVID-19, whereas negative indicates that the patient does not have COVID-19. |
| 2 | Conformation-Method | The Conformation-Method can be either Clinical or RT-PCR. Clinical is based on doctor observations. RT-PCR is a COVID-19 test. |
| 3 | Age | The age of patient. |
| 4 | Ventilator | A device that produces or assists pulmonary ventilation.<br>This feature shows whether the patient needs a Ventilator or not. |
| 5 | Cough | During this process, air is expelled quickly through a partially closed glottis with an audible sound.<br>Determines whether the patient has a cough or not. |
| 6 | Apnea | An absence of spontaneous breathing. |
| 7 | Carcinoma | It is a type of cancer caused by epithelial cells found in the lining of various organs and tissues. |
| 8 | Healthcare_staff | Personnel working in the health care industry. |
| 9 | ICU_hospitalization | Admission of a patient to the ICU for close monitoring and specialized care. |

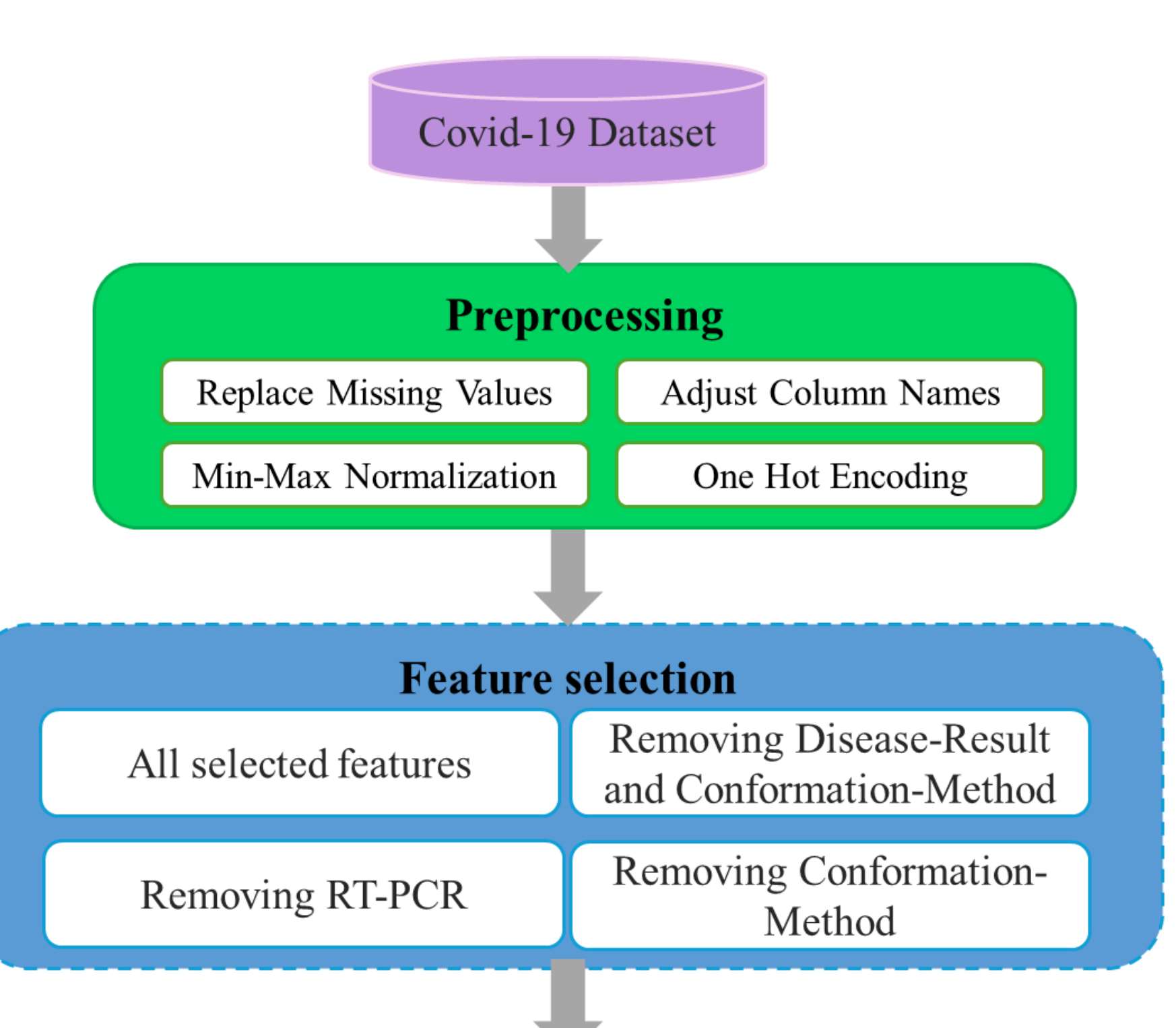
Covid-19 Dataset
Preprocessing
Replace Missing Values
Adjust Column Names
Min-Max Normalization
One Hot Encoding
Feature selection
All selected features
Removing Disease-Result and Conformation-Method
Removing RT-PCR
Removing Conformation-Method

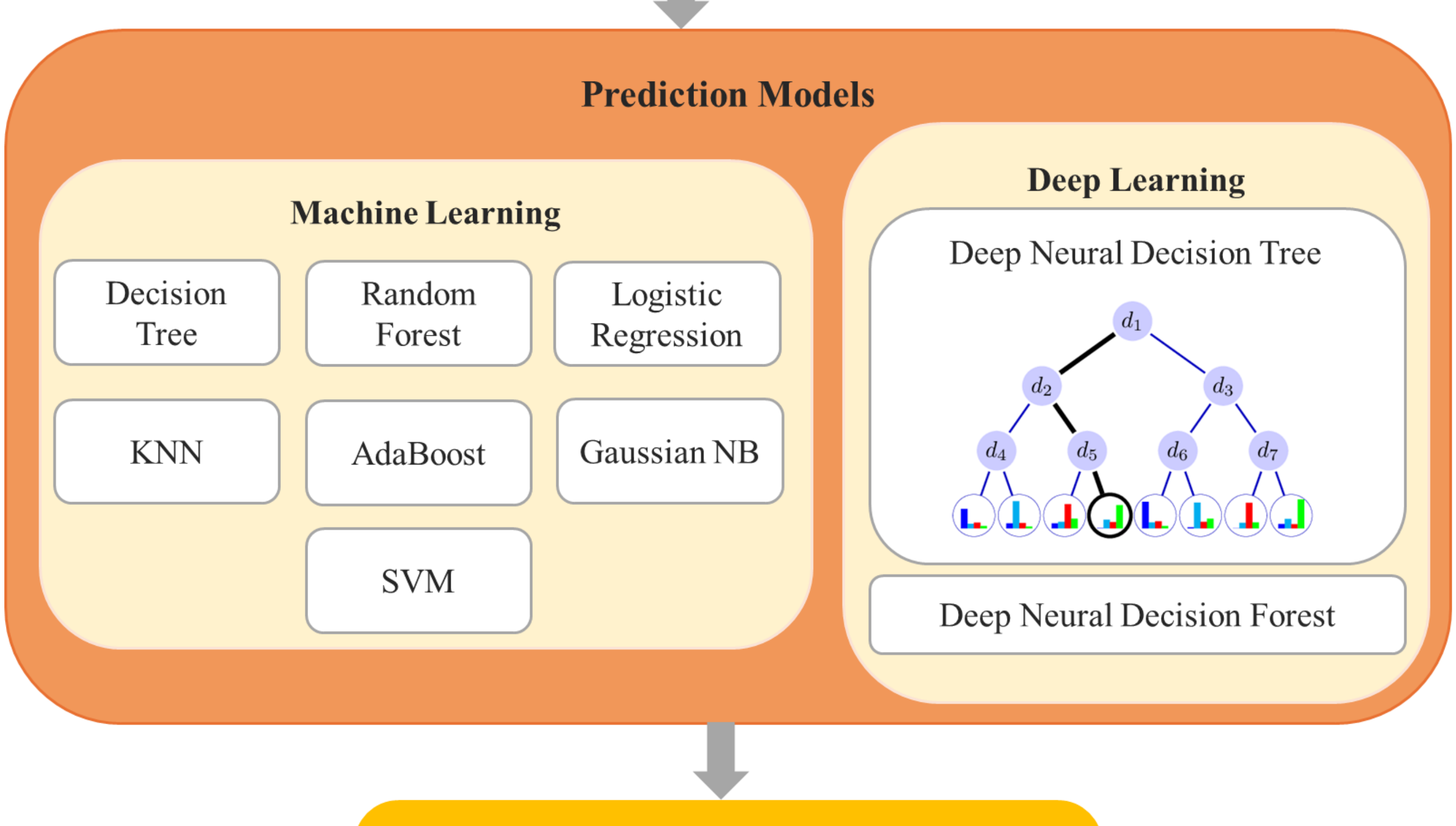
Prediction Models
Machine Learning
Decision Tree
Random Forest
Logistic Regression
KNN
AdaBoost
Gaussian NB
SVM
Deep Learning
Deep Neural Decision Tree
$d_1$
$d_2$
$d_3$
$d_4$
$d_5$
$d_6$
$d_7$
Deep Neural Decision Forest

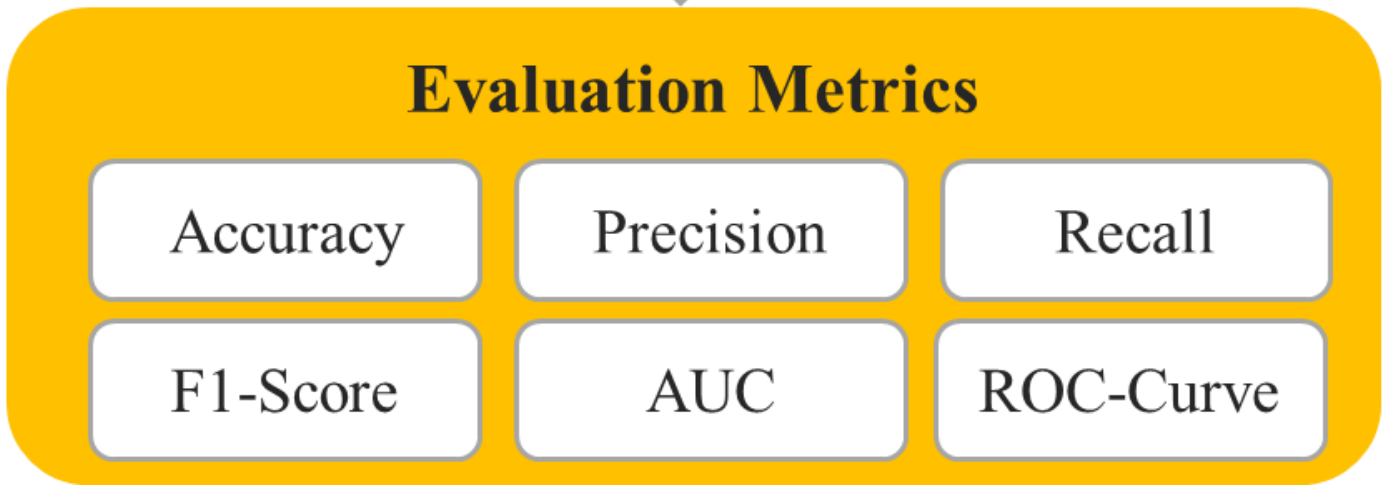
Evaluation Metrics
Accuracy
Precision
Recall
F1-Score
AUC
ROC-Curve

Figure 1: Proposed model.

We investigated the skewness and kurtosis values of the features (Table 2). The age feature is approximately symmetric and normally distributed, while the remaining features exhibit varying degrees of skewness and kurtosis, indicating non-normal distributions. Specifically, Ventilator, Carcinoma, Healthcare-staff and ICU-hospitalization are highly skewed and heavy-tailed, while Cough and Apnea are moderately skewed and light-tailed. The correlation between features also depicted in Figure 2. The higher positive correlations were found bewteen Test-Result and Conformation-Method (r = 1.0) . Due to this complete multicollinearity, we excluded this feature from multiple phases of our analysis to avoid redundancy and potential statistical issues. The negative correlation between age and health-care staff (r=-0.2).

Table 2. The skewness and kurtosis values of the features.

| Features | Skewness | Skewness Interpretation | Kurtosis | Kurtosis Interpretation |
|---|---|---|---|---|
| Age | -0.387 | approximately symmetric | -0.359 | approximately normal |
| Ventilator | 1.628 | highly skewed | 0.651 | heavy-tailed (leptokurtic) |
| Cough | -0.577 | moderately skewed | -1.667 | light-tailed (platykurtic) |
| Apnea | -0.97 | moderately skewed | -1.058 | light-tailed (platykurtic) |
| Carcinoma | 6.269 | highly skewed | 37.304 | heavy-tailed (leptokurtic) |
| Healthcare-staff | 3.563 | highly skewed | 10.693 | heavy-tailed (leptokurtic) |
| ICU-hospitalization | 2.422 | highly skewed | 3.868 | heavy-tailed (leptokurtic) |

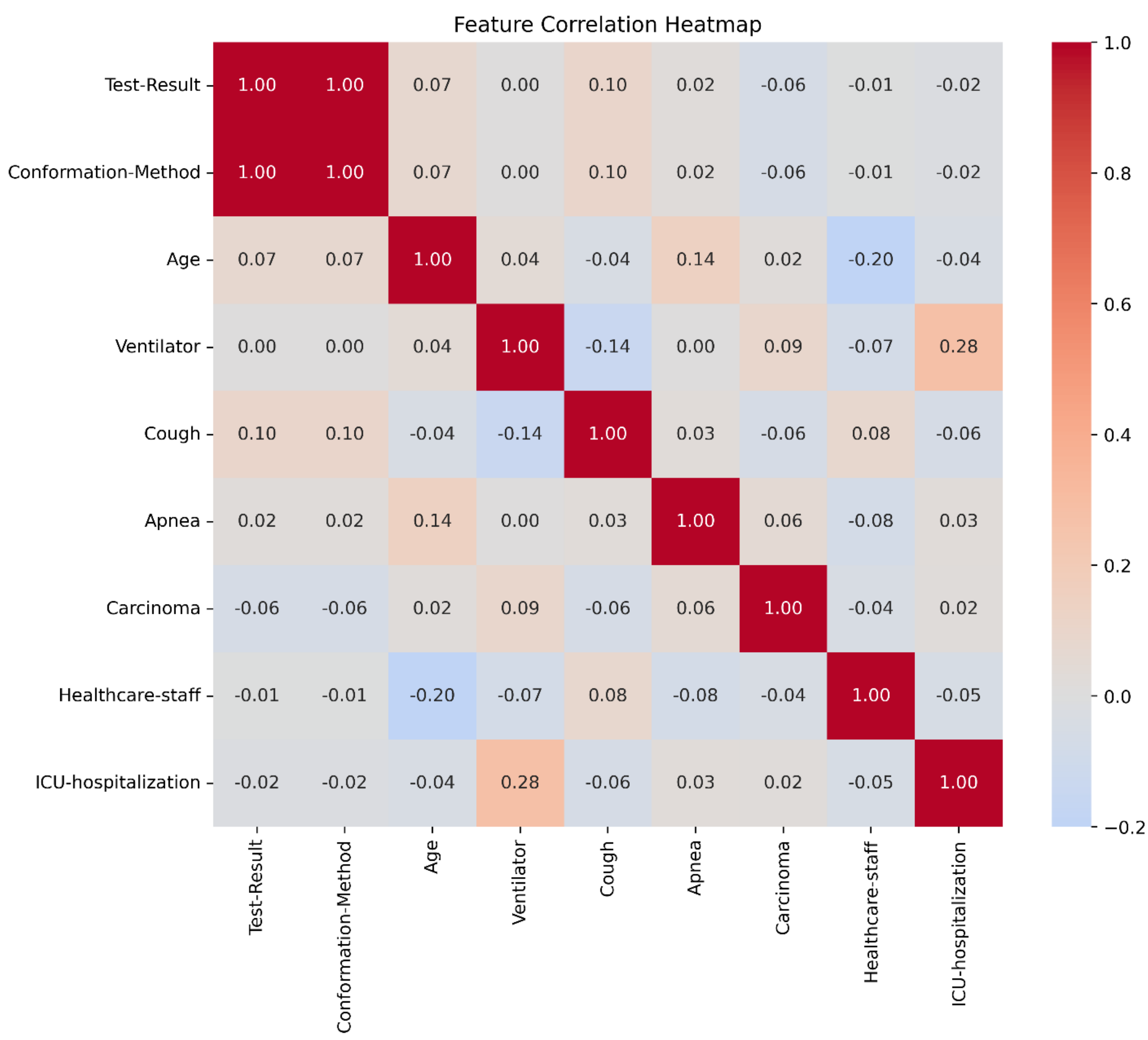

Figure 2. Heatmap indicating the correlation between selected feature.

A stratified sampling method was employed to partition the dataset, allocating 80% for training and 20% for testing the models. Stratified sampling involves selecting samples in proportions that reflect their distribution in the overall population by dividing it into groups, or "strata," based on specific characteristics (2). This approach enhances the flexibility of sampling methods across different strata and provides more accurate estimates of target parameters.

The implementation was carried out in four distinct stages. In the first stage, all features listed in Table 1 were included when training the model. In the second stage, we removed the Test-Result and Conformation-Method features from the dataset and built models using the remaining features. The Conformation-Method feature contains two values: Clinical and RT-PCR. In the third stage, we focused solely on the Clinical data, excluding RT-PCR from the dataset. Finally, in the fourth stage, we used only the RT-PCR data for predictions. After preparing the data, we applied seven machine learning methods and two deep learning techniques to train the model for predicting patient recovery or mortality at each stage. The popular traditional machine learning including Decision Tree, Random Forest, Logistic Regression, K-Nearest Neighbors (KNN), Adaptive Boosting (AdaBoost), Gaussian Naive Bayes (Gaussian NB), Support Vector Machine (SVM) (19), and some deep learning algorithms including Deep Neural Decision Tree (20) and Deep Neural Decision Forest (21) were used to build classifiers which are explained in the following. Traditional machine learning models face several limitations, including dependence on data quality and manual feature engineering, limited interpretability, and challenges handling unstructured, sequential, and missing data (22, 23).

As illustrated in Figure 3, the proposed deep neural decision tree consists of several layers, starting with the input layer, which includes selected features organized into four stages; in stages one, three, and four, there are nine input features, while the second stage contains seven features. Following the input layer, multiple hidden layers are added to identify and analyze complex patterns within the data, incorporating processes such as dimensional expansion, concatenation, and batch normalization. This structure allows for a random subset of input features to be selected, after which probabilities are computed for the input instances to reach the tree leaves. Finally, the output layer is represented by the neural decision tree, which produces the final results using a Sigmoid activation function.

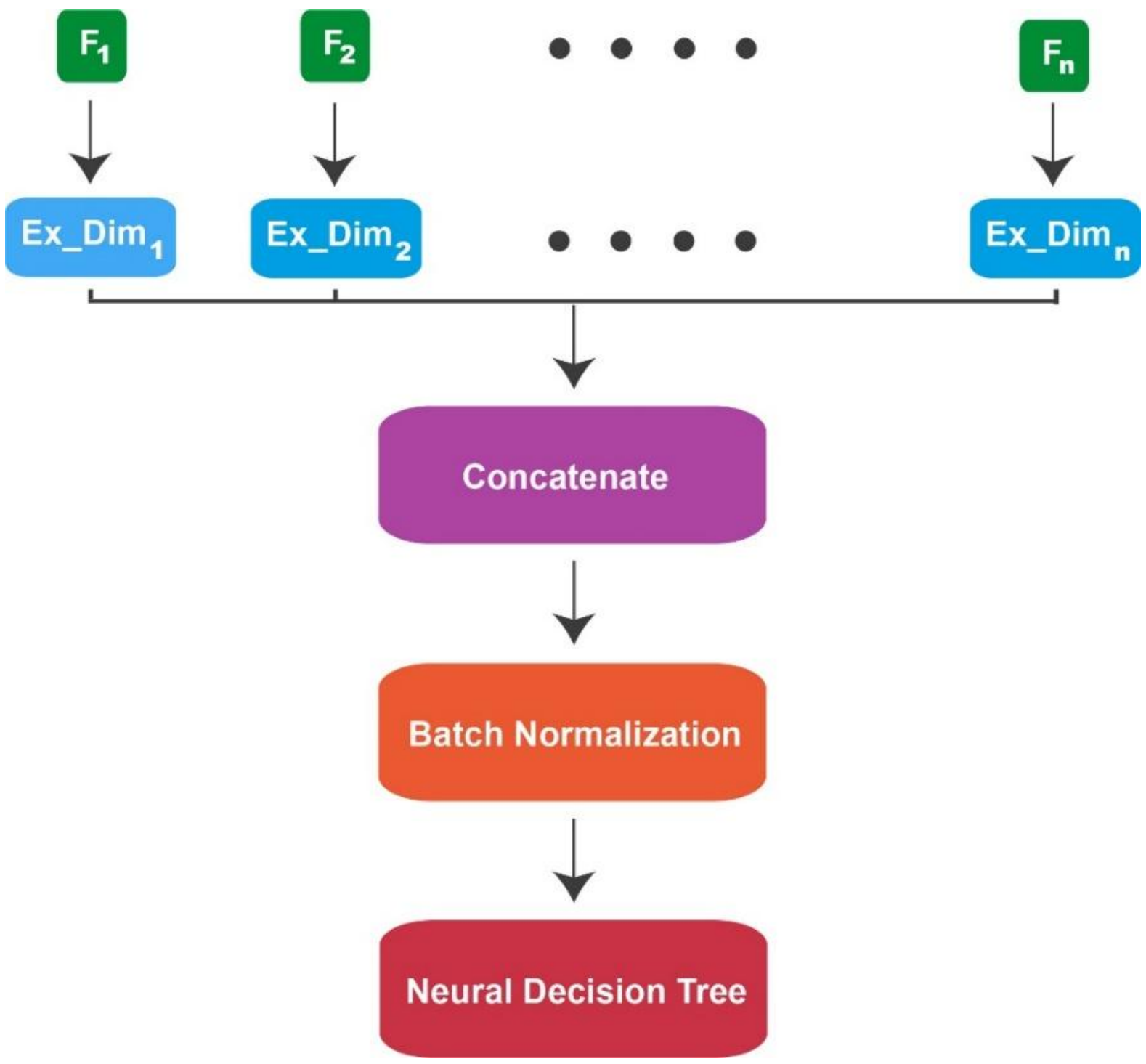

Figure 3: Proposed deep neural decision tree architecture (F: Feature, Ex_Dim, Expand Dimension).

Deep neural decision forest is a model that combines the strengths of decision forests with the non-linear transformations learned by deep neural networks (24). Instead of a single decision tree, deep neural decision forest uses an ensemble of decision trees to make predictions (25). The combination of deep learning with the interpretability and ensemble nature of decision forest models, makes it possible to extract complex patterns from data (21). In neural decision forests, several neural decision trees are simultaneously trained. In a forest model, the output is the average output of the trees. As shown in Table 3, we utilized 25 trees with depth of 10, and the hyperparameters equal to 16, 30, and 0.001 for the batch size, epochs, and learning rate. Moreover, the Adam optimization algorithm was used and the loss function was Binary Cross Entropy. We optimized these parameters through automated hyperparameter tuning using Ray Tune, a Python-based distributed framework that efficiently searches the parameter space to identify optimal values.

Table 3. The deep neural decision forest parameters.

| Parameter | Value |
|---|---|
| Number of trees | 25 |
| Depth | 10 |
| Batch size | 16 |
| Epochs | 30 |
| Learning rate | 0.001 |
| Optimizer | Adam |
| Loss function | Binary Cross Entropy |

## Results

### Dataset

In the dataset, 2875 samples were collected from March 2020 to March 2021 of Mashhad hospitals (same as (1) dataset). A total of 1787 Clinical samples and 1088 RT-PCR samples have been collected. Figure 4a illustrates the recovery/death of patients based on their Test-Result feature. There is no surprise that those with a negative result have a higher recovery rate which is about 85%. Approximately 75% of patients with positive results recovered, and sadly, 25% died. Figure 4b shows the number of recovered and deceased patients based on their age. With increasing age, there is a greater possibility of death. One interesting point is that all patients under the age of 40 recovered from the disease. It should be noted that this analysis is based on this research dataset and may differ in other datasets.

Figure 4c shows the percentage of patients who recovered or died based on their gender. Men's results are represented by the zero (blue color) and women's results are represented by the one (orange color). The death risk among younger men is higher than among younger women. In Figure 4d, it can be seen that the sooner a patient is discharged from hospital, the greater the chance of survival.

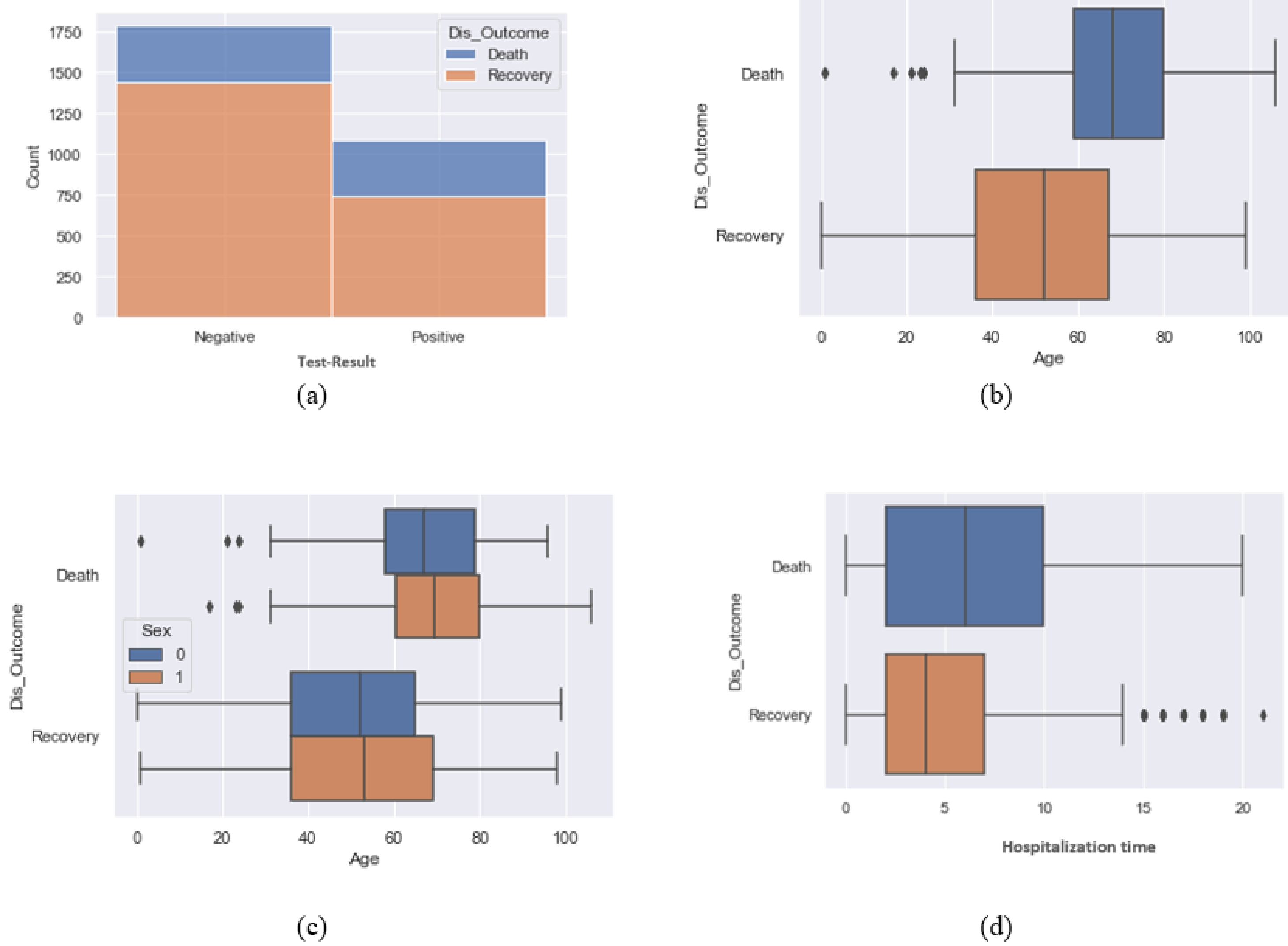

Figure 4: The statics of dataset: (a) Number of recovery or death of patients, (b) Number of recovered and deceased patients based on their age, (c) Number of recovered and deceased patients based on their gender, (d) The effect of patient's hospitalization time on her/his survival.

**Comparative evaluation of mortality prediction**

This section provides the results and a comparative evaluation of mortality prediction in COVID-19 patients, assessing the effectiveness of clinical assessments and RT-PCR testing. The results of the first stage are shown in Table 4. The deep neural decision forest model outperforms all the other used models with an accuracy and recall of 78.3%, and F1-score of 74.1%. The deep neural decision tree model is the second-best algorithm. In this method, the accuracy, recall and precision are equal to 77.4% and F1-score is equal to 74%. Among the machine learning methods, random forest achieved the best results, with accuracy and recall of 76.9%, precision of 73.7%, and F1-score of 73.8%. AdaBoost generated the closest results to random forest. Ensemble learning is an important characteristic of these algorithms.

Table 4. First stage results.

| Model name | Accuracy | Recall | Precision | F1-score |
|---|---|---|---|---|
| Gaussian NB | 0.563 | 0.563 | 0.785 | 0.586 |
| KNN | 0.751 | 0.751 | 0.719 | 0.727 |
| Logistic Regression | 0.769 | 0.769 | 0.733 | 0.725 |
| Decision Tree | 0.689 | 0.689 | 0.688 | 0.688 |
| Random Forest | 0.769 | 0.769 | 0.737 | 0.738 |
| SVM | 0.758 | 0.758 | 0.575 | 0.654 |
| AdaBoost | 0.765 | 0.765 | 0.735 | 0.739 |
| Deep Neural Decision Tree | 0.774 | 0.774 | 0.744 | 0.740 |
| Deep Neural Decision Forest | 0.783 | 0.783 | 0.758 | 0.741 |

Figure 5(a,b) shows the confusion matrix and classification report for the deep neural decision forest model at the first stage. Our model correctly predicted the recovery of 418 and the death of 32 patients. The model misclassified 125 patients. In general, the majority of samples are classified correctly, but the false positive rate is high. In medical domain, higher false positive is more acceptable than higher false negative, because it doesn't put the patient in serious dangerous.

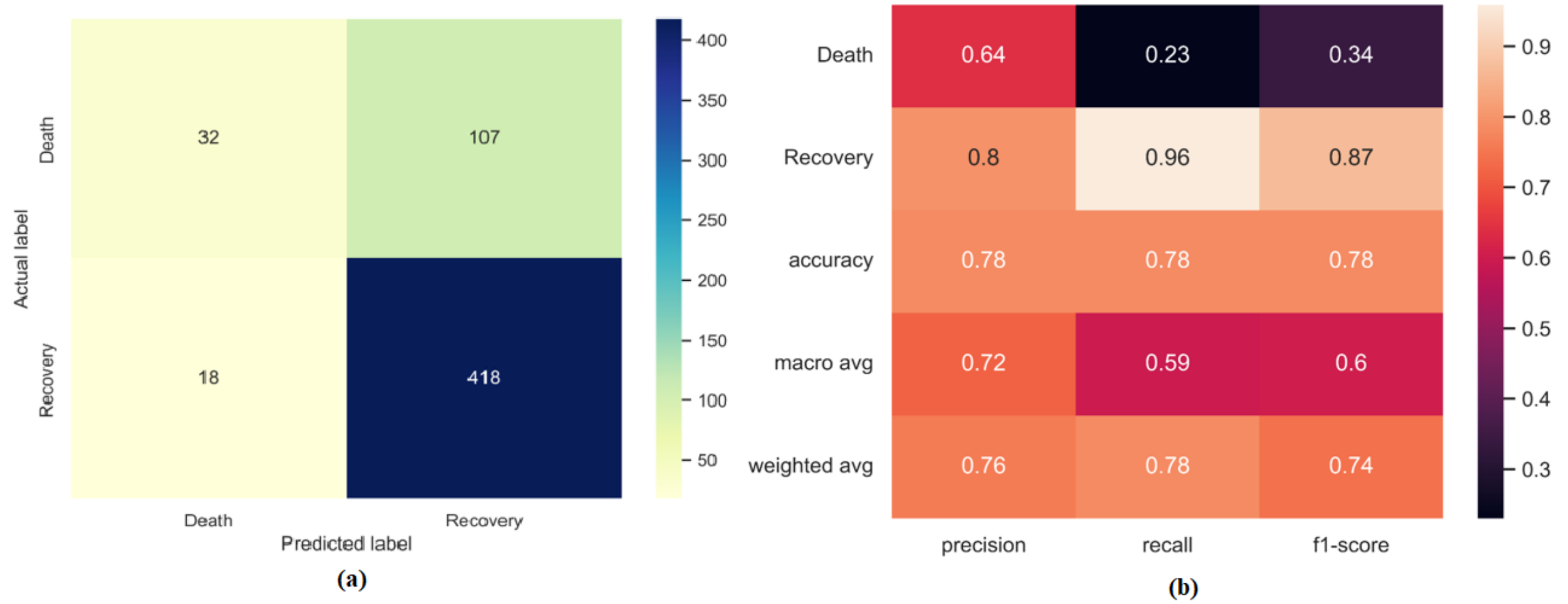

Figure 5: (a) Confusion matrix and (b) classification report of the first stage.

Table 5 presents the results of the second stage. With an accuracy and recall of 77.9% and an F1-score of 75.3%, the deep neural decision forest model performs better than the other models. The results are not very different from the first stage, indicating that the removal of Test-Result and Conformation-Method features does not significantly affect the performance of the model.

Table 5. Second stage results.

| Model name | Accuracy | Recall | Precision | F1-score |
|---|---|---|---|---|
| Gaussian NB | 0.570 | 0.570 | 0.804 | 0.592 |
| KNN | 0.753 | 0.753 | 0.721 | 0.728 |
| Logistic Regression | 0.772 | 0.772 | 0.740 | 0.726 |
| Decision Tree | 0.701 | 0.701 | 0.716 | 0.707 |
| Random Forest | 0.770 | 0.770 | 0.737 | 0.732 |
| SVM | 0.758 | 0.758 | 0.575 | 0.654 |
| AdaBoost | 0.760 | 0.760 | 0.738 | 0.661 |

| Deep Neural Decision Tree | 0.777 | 0.777 | 0.749 | 0.740 |
|---|---|---|---|---|
| Deep Neural Decision Forest | 0.779 | 0.779 | 0.753 | 0.753 |

Based on Figure 6 (a,b), there is a correct prediction of recovery for 405 patients and an incorrect prediction for 31 patients. In addition, 43 deaths have been predicted correctly and 96 have been classified incorrectly.

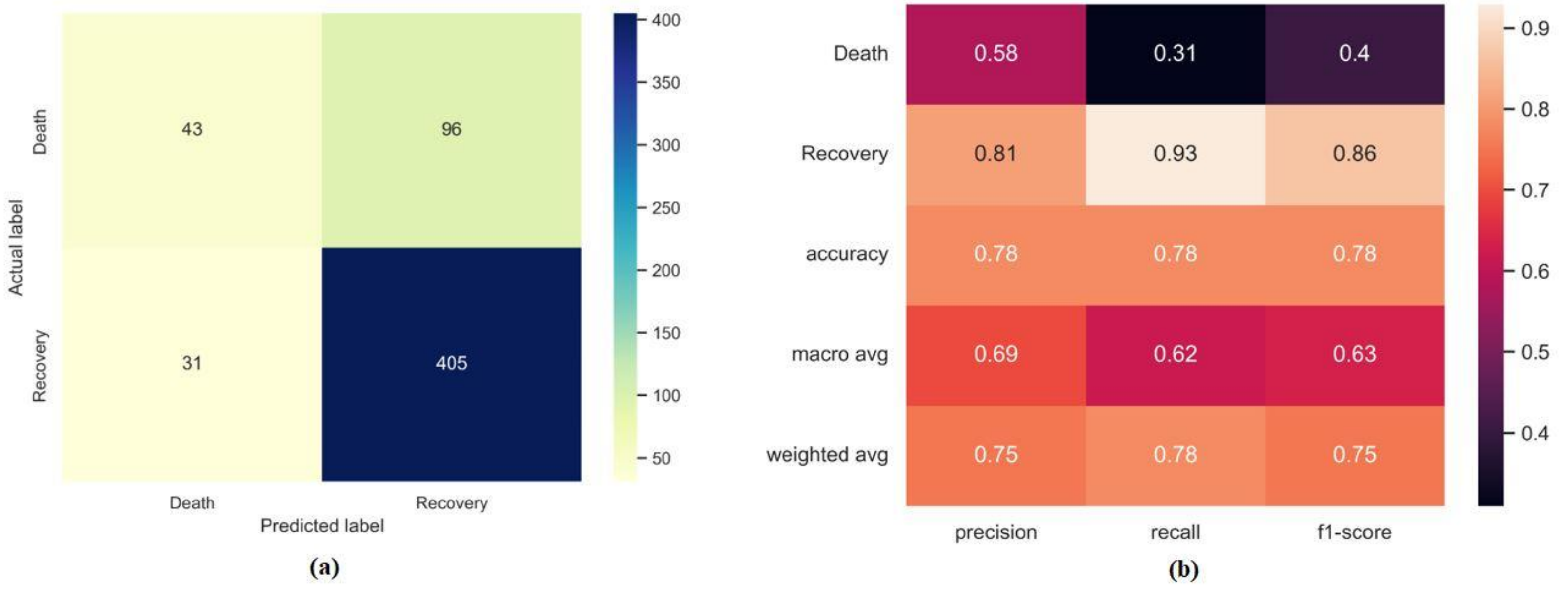

Figure 6: (a) Confusion matrix and (b) classification report of the second stage.

Table 6 presents the results of the models in the third stage, showing that all models outperform those from the first. Notably, the deep neural decision forest remains the most effective model overall, with an accuracy increase of 2.4% compared to the first stage. This suggests that clinical assessments positively influence predictions regarding recovery or mortality. Furthermore, the removal of RT-PCR results from the dataset appears to enhance the performance of all models, indicating that they are sensitive to false negatives.

.

Table 6. Third stage results.

| Model name | Accuracy | Recall | Precision | F1-score |
|---|---|---|---|---|
| Gaussian NB | 0.257 | 0.257 | 0.847 | 0.185 |
| KNN | 0.804 | 0.804 | 0.651 | 0.720 |
| Logistic Regression | 0.807 | 0.807 | 0.751 | 0.731 |
| Decision Tree | 0.718 | 0.718 | 0.729 | 0.723 |
| Random Forest | 0.807 | 0.807 | 0.652 | 0.721 |
| SVM | 0.807 | 0.807 | 0.652 | 0.721 |
| AdaBoost | 0.788 | 0.788 | 0.731 | 0.746 |
| Deep Neural Decision Tree | 0.804 | 0.804 | 0.751 | 0.750 |
| Deep Neural Decision Forest | 0.807 | 0.807 | 0.757 | 0.748 |

This stage's confusion matrix is shown in Figure 7a. Out of 358 test cases, 289 were recovered in which the model correctly identified 283 and had an error for only 6 cases. Therefore, the false negative rate is very low, which indicate clinical features are appropriate for decision making (Figure 7b).

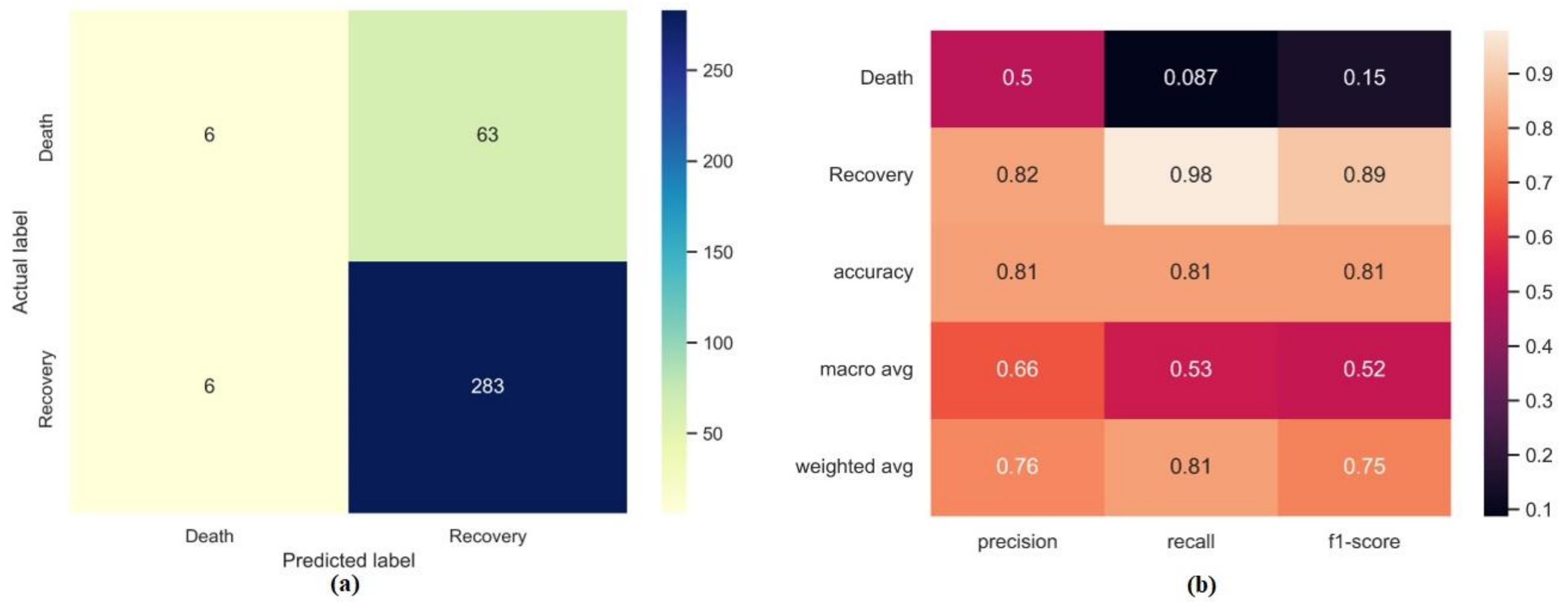

Figure 7: (a) Confusion matrix and (b) classification report of the third stage.

The results obtained at the fourth stage are presented in Table 7. The deep neural decision forest model still has the best results, but the performance of all models has decreased. RT-PCR negatively affects the performance of all models and reduces the accuracy of the deep neural decision forest by 9%, 8.6%, and 11.4% compared to the first, second, and third stages, respectively. This is because RT-PCR has a high false-positives and false-negatives rate.

Table 7. Fourth stage results.

| Model name | Accuracy | Recall | Precision | F1-score |
|---|---|---|---|---|
| Gaussian NB | 0.560 | 0.560 | 0.710 | 0.562 |
| KNN | 0.656 | 0.656 | 0.630 | 0.637 |
| Logistic Regression | 0.679 | 0.679 | 0.642 | 0.637 |
| Decision Tree | 0.583 | 0.583 | 0.597 | 0.589 |
| Random Forest | 0.670 | 0.670 | 0.620 | 0.611 |
| SVM | 0.679 | 0.679 | 0.461 | 0.549 |
| AdaBoost | 0.642 | 0.642 | 0.624 | 0.630 |
| Deep Neural Decision Tree | 0.693 | 0.693 | 0.667 | 0.666 |
| Deep Neural Decision Forest | 0.693 | 0.693 | 0.674 | 0.678 |

According to Figure 8a, the confusion matrix shows the deep neural decision forest model has the total 151 true predictions and 67 false predictions. The classification metrics also specified in Figure 8b.

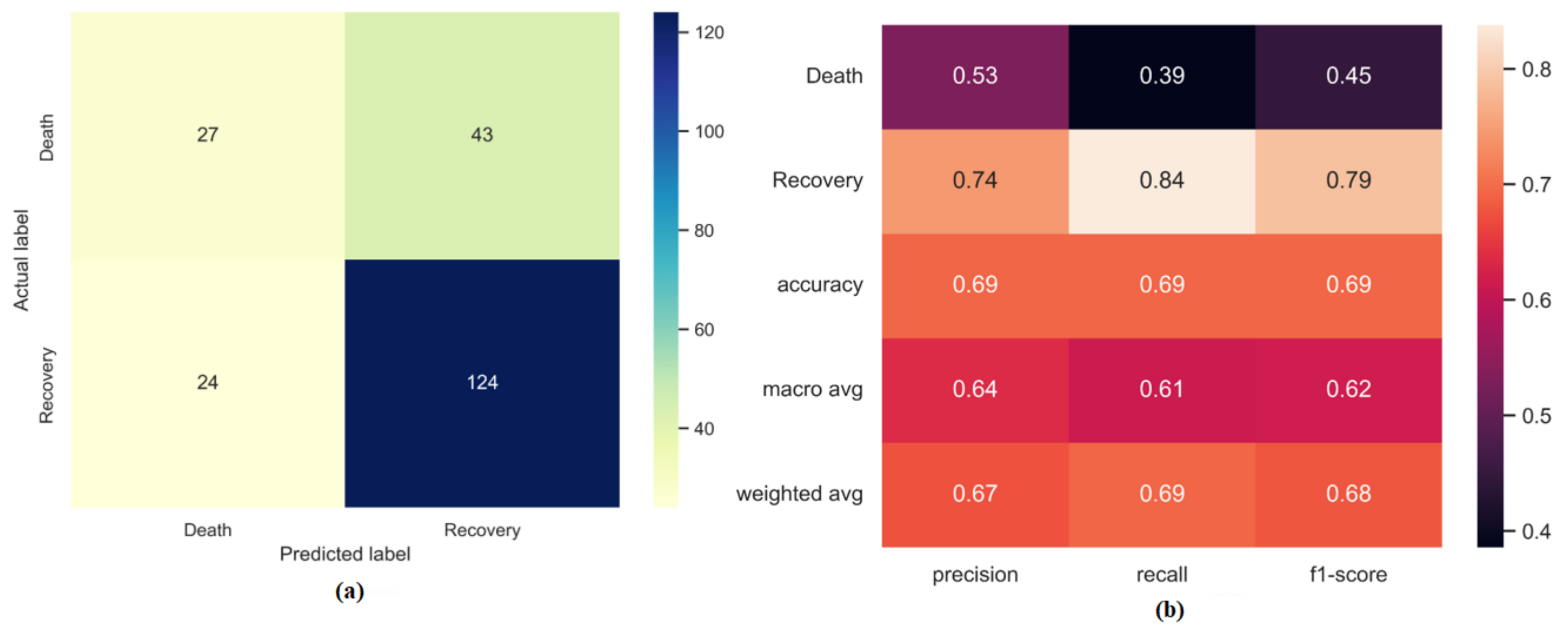

Figure 8: (a) Confusion matrix and (b) classification report of the fourth stage.

To comprehensively evaluate model performance, we generated Receiver Operating Characteristic (ROC) curves for each deep learning model across all four stages, as shown in Figure 9. Among all models and stages, the deep neural decision forest achieved the highest Area Under the Curve (AUC) of 0.782 during stage 3.

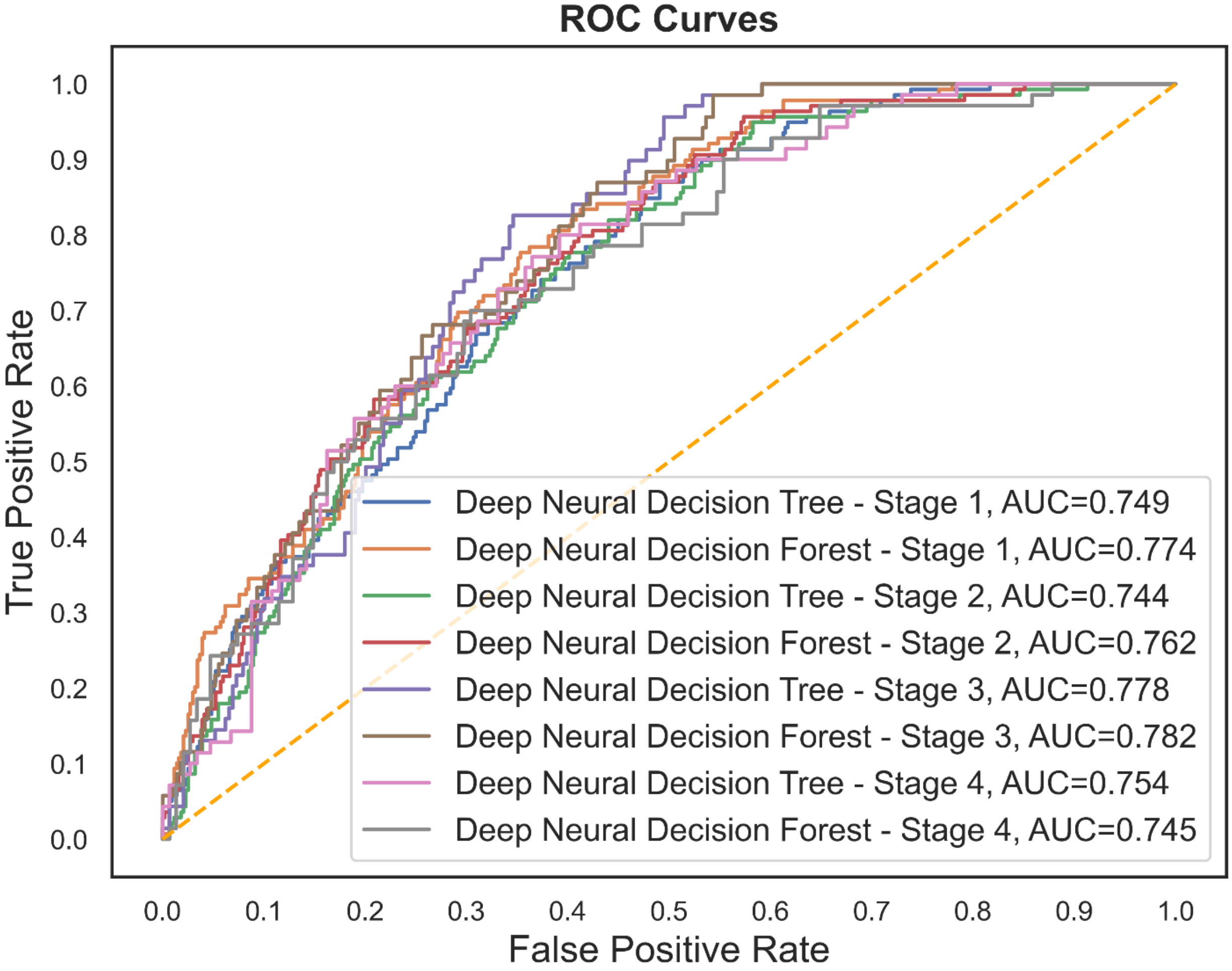

Figure 9. ROC curves of deep neural decision forest and deep neural decision tree models for all four stages.

To assess the performance of our proposed models (Deep Neural Decision Tree and Deep Neural Decision Forest), we conducted a comparative analysis against a previously published method addressing a similar task (16). We fully implemented the benchmark method and evaluated it on our dataset. The comparison, presented in Table 8, was based on three key metrics: AUC, micro-averaged sensitivity, and micro-averaged specificity.

The Deep Neural Decision Forest achieved its optimal performance during stage 3, with an AUC of 0.782, micro-averaged sensitivity of 0.81, and micro-averaged specificity of 0.81. Our method outperformed the results obtained using the artificial neural network (ANN) approach described in ref (16) for predicting COVID-19 mortality when both were evaluated on our dataset.

Table 8. Comparison of models.

| Stage | Model | AUC | Sensitivity | Specificity |
|---|---|---|---|---|
| 1 | Deep Neural Decision Tree | 0.749 | 0.767 | 0.767 |
| | Deep Neural Decision Forest | 0.774 | 0.777 | 0.777 |
| | (16) | 0.732 | 0.748 | 0.748 |
| 2 | Deep Neural Decision Tree | 0.744 | 0.758 | 0.758 |
| | Deep Neural Decision Forest | 0.762 | 0.769 | 0.769 |
| | (16) | 0.721 | 0.743 | 0.743 |
| 3 | Deep Neural Decision Tree | 0.778 | 0.799 | 0.799 |
| | Deep Neural Decision Forest | **0.782** | **0.81** | **0.81** |
| | (16) | 0.767 | 0.793 | 0.793 |
| 4 | Deep Neural Decision Tree | 0.754 | 0.711 | 0.711 |
| | Deep Neural Decision Forest | 0.745 | 0.702 | 0.702 |
| | (16) | 0.733 | 0.725 | 0.725 |

## Discussion

Artificial intelligence (AI) is revolutionizing healthcare through improved predictive analytics, optimized decision-making, and enhanced patient outcomes (26). AI models, including deep neural models, are increasingly used for disease prognosis due to their ability to analyze large clinical datasets accurately. However, challenges related to data bias, model interpretability, and computational costs require careful consideration (27).

Predicting patient recovery or mortality is crucial during disease outbreaks, particularly when medical resources are scarce. This study evaluated the effectiveness of machine learning and specifically deep neural decision forest and tree models in predicting outcomes for COVID-19 patients as data model across four stages, focusing on the roles of clinical data. This highlights the potential of clinical assessments as reliable predictors of patient outcomes, even surpassing the predictive capabilities of RT-PCR testing.

As medical diagnosis has progressed, nucleic acid detection-based methods have become an essential tool for detecting viral infections. COVID-19 has prompted the development of more than seven types of nucleic acid testing kits (28). Since RT-PCR is a specific and simple qualitative test, it has gained increasing popularity (29). However, RT-PCR kits are insufficient, experts are in high demand, false negative rates are high, accredited laboratories are required, infrastructure is costly, and qualified workers are necessary (30). Moreover, a negative result does not necessarily indicate the absence of COVID-19 infection, and should not be relied upon as an absolute criterion for treatment (28).

Recent studies on COVID-19 mortality prediction have leveraged advanced machine learning techniques and imaging data. One study utilized 201 radiomic features and 16 neural network features from chest X-rays of 1,816 patients, applying algorithms like AdaBoost and random forest to forecast mortality (31). Another research employed convolutional neural networks on data from 1,170 patients with 19,685 labeled CT slices to predict outcomes. Additionally, a multivariable logistic regression model analyzed risk factors in 1,663 hospitalized patients, identifying key variables such as age and pre-existing medical conditions that influence mortality rates (32). Consistent with theses study, we found that the deep neural decision forest achieved an accuracy of 80.7%, demonstrating advantages such as reduced overfitting and improved robustness to noise. The deep neural decision forest also performed well, with accuracy of 80.4%. Therefore, the proposed approches in this reserach demonstrate the potential of artificial intelligence to effectively predict patient recovery or mortality by utilizing a variety of relevant clinical features. Furthermore, it emphasizes the relative importance of clinical data over RT-PCR results in predicting outcomes.

Future research directions should address several key areas to enhance the model's effectiveness. The model's generalizability could be improved by incorporating more diverse and representative patient populations, thereby addressing potential biases in the data sources. The model's performance could potentially be enhanced by integrating additional data modalities, particularly chest imaging data, which has demonstrated significant diagnostic value in related studies.

## Conclusion

This study demonstrates the utility of deep neural decision forest and deep neural decision tree models for predicting mortality among COVID-19 patients, addressing the critical need for effective risk stratification, particularly in resource-limited settings. The deep neural decision forest outperformed other models across various stages of analysis when using only clinical data. This suggests its potential as a reliable and accessible tool for predicting patient mortality. Moreover, the model's ability to learn complex patterns from clinical data while offering some interpretability is crucial for clinical adoption, enabling healthcare professionals to understand the factors driving predictions. Its strong performance with readily available clinical data reduces reliance on costly or invasive tests, improving accessibility. However, the deep neural decision forest misclassifies some patients and has a high false positive rate, although this is often acceptable in medical settings.

## Abbreviations

CT: computed tomography

RT-PCR: Reverse transcription polymerase chain reaction

SVM: Support Vector Machine

Gaussian NB: Gaussian Naive Bayes

KNN: K-Nearest Neighbors

AdaBoost: Adaptive Boosting

## Declarations

### Ethics approval and consent to participate

Not applicable.

### Consent for publication

Not applicable.

### Availability of data and materials

All data generated or analysed during this study are included in this published article [and its supplementary information files].

## Competing interests

The authors declare that they have no competing interests.

## Funding

Not applicable.

## Authors' contributions

Conceptualization, M. D.; Formal analysis, M. D. and D. T. D.; Methodology, M. D.; Software, M. D.; Supervision, M. D.; Validation, M. D. and M-ZG; Visualization, M. D.; Writing – original draft, M. M.; Writing – review & editing, M. D., M-ZG, and D. T. D. All authors have read and agreed to the published version of the manuscript.

## Acknowledgements

Not applicable.

## References

1. Dehghani M, Yazdanparast Z. Discovering the symptom patterns of COVID-19 from recovered and deceased patients using Apriori association rule mining. Informatics in Medicine Unlocked. 2023;42:101351.
2. Parsons VL. Stratified sampling. Wiley StatsRef: Statistics Reference Online. 2014:1-11.
3. Pruszyński J, Cianciara D, Pruszyńska I, Włodarczyk-Pruszyńska I. Staff shortages and inappropriate work conditions as a challenge geriatrics and contemporary healthcare service at large faces. Journal of Education, Health and Sport. 2022;12(7):136-47.
4. Ahmad RW, Salah K, Jayaraman R, Yaqoob I, Ellahham S, Omar M. Blockchain and COVID-19 pandemic: Applications and challenges. Cluster Computing. 2023;26(4):2383-408.
5. Adekugbe AP, Ibeh CV. Harnessing data insights for crisis management in us public health: lessons learned and future directions. International Medical Science Research Journal. 2024;4(4):391-405.
6. Gao J, Lu Y, Ashrafi N, Domingo I, Alaei K, Pishgar M. Prediction of Sepsis Mortality in ICU patients using machine learning methods. BMC Medical Informatics and Decision Making. 2024;24(1):228.
7. Liu M, Guo C, Guo S. An explainable knowledge distillation method with XGBoost for ICU mortality prediction. Computers in Biology and Medicine. 2023;152:106466.

8. Desai RJ, Wang SV, Vaduganathan M, Evers T, Schneeweiss S. Comparison of machine learning methods with traditional models for use of administrative claims with electronic medical records to predict heart failure outcomes. JAMA network open. 2020;3(1):e1918962-e.
9. Goh KH, Wang L, Yeow AYK, Poh H, Li K, Yeow JJL, et al. Artificial intelligence in sepsis early prediction and diagnosis using unstructured data in healthcare. Nature communications. 2021;12(1):711.
10. Mainali S, Darsie ME, Smetana KS. Machine learning in action: stroke diagnosis and outcome prediction. Frontiers in neurology. 2021;12:734345.
11. Song X, Liu X, Liu F, Wang C. Comparison of machine learning and logistic regression models in predicting acute kidney injury: A systematic review and meta-analysis. International journal of medical informatics. 2021;151:104484.
12. Pourhomayoun M, Shakibi M. Predicting mortality risk in patients with COVID-19 using machine learning to help medical decision-making. Smart health. 2021;20:100178.
13. Waller JV, Kaur P, Tucker A, Lin KK, Diaz MJ, Henry TS, et al. Diagnostic tools for coronavirus disease (COVID-19): comparing CT and RT-PCR viral nucleic acid testing. American Journal of Roentgenology. 2020;215(4):834-8.
14. Al-Momani H. A Literature Review on the Relative Diagnostic Accuracy of Chest CT Scans versus RT-PCR Testing for COVID-19 Diagnosis. Tomography [Internet]. 2024; 10(6):[935-48 pp.].
15. Queipo M, Barbado J, Torres AM, Mateo J. Approaching personalized medicine: the use of machine learning to determine predictors of mortality in a population with SARS-CoV-2 infection. Biomedicines. 2024;12(2):409.
16. Talkhi N, Akbari Sharak N, Yousefi R, Salari M, Sadati SM, Shakeri MT. Predicting COVID-19 Mortality and Identifying Clinical Symptom Patterns in Hospitalized Patients: A Machine-learning Study. Iranian Journal of Health Sciences. 2024;12(1):39-48.
17. Álvarez-Esteban PC, Del Barrio E, Rueda OM, Rueda C. Predicting COVID-19 progression from diagnosis to recovery or death linking primary care and hospital records in Castilla y León (Spain). PLoS One. 2021;16(9):e0257613.
18. Ortiz A, Trivedi A, Desbiens J, Blazes M, Robinson C, Gupta S, et al. Effective deep learning approaches for predicting COVID-19 outcomes from chest computed tomography volumes. Scientific reports. 2022;12(1):1716.
19. Afsaneh E, Sharifdini A, Ghazzaghi H, Ghobadi MZ. Recent applications of machine learning and deep learning models in the prediction, diagnosis, and management of diabetes: a comprehensive review. Diabetology & metabolic syndrome. 2022;14(1):196.
20. Santos RS, Festijo ED, editors. Classifying Portable Executable Malware Using Deep Neural Decision Tree2022: IEEE.
21. Alrayes FS, Zakariah M, Driss M, Boulila W. Deep Neural Decision Forest (DNDF): A Novel Approach for Enhancing Intrusion Detection Systems in Network Traffic Analysis. Sensors. 2023;23(20):8362.
22. Barbierato E, Gatti A. The Challenges of Machine Learning: A Critical Review. Electronics [Internet]. 2024; 13(2).
23. Ezugwu AE, Ho Y-S, Egwuche OS, Ekundayo OS, Van Der Merwe A, Saha AK, et al. Classical Machine Learning: Seventy Years of Algorithmic Learning Evolution. arXiv preprint arXiv:240801747. 2024.
24. Zhou Z-H, Feng J. Deep forest. National science review. 2019;6(1):74-86.
25. Kontschieder P, Fiterau M, Criminisi A, Bulo SR, editors. Deep neural decision forests. Proceedings of the IEEE international conference on computer vision; 2015.
26. Alowais SA, Alghamdi SS, Alsuhebany N, Alqahtani T, Alshaya AI, Almohareb SN, et al. Revolutionizing healthcare: the role of artificial intelligence in clinical practice. BMC medical education. 2023;23(1):689.

27. Minh D, Wang HX, Li YF, Nguyen TN. Explainable artificial intelligence: a comprehensive review. Artificial Intelligence Review. 2022:1-66.
28. Wang Y, Kang H, Liu X, Tong Z. Combination of RT‐qPCR testing and clinical features for diagnosis of COVID‐19 facilitates management of SARS‐CoV‐2 outbreak. Journal of medical virology. 2020;92(6):538.
29. Shen M, Zhou Y, Ye J, Al-Maskri AAA, Kang Y, Zeng S, et al. Recent advances and perspectives of nucleic acid detection for coronavirus. Journal of pharmaceutical analysis. 2020;10(2):97-101.
30. Abayomi-Alli OO, Damaševičius R, Maskeliūnas R, Misra S. An ensemble learning model for COVID-19 detection from blood test samples. Sensors. 2022;22(6):2224.
31. Iori M, Di Castelnuovo C, Verzellesi L, Meglioli G, Lippolis DG, Nitrosi A, et al. Mortality prediction of COVID-19 patients using radiomic and neural network features extracted from a wide chest X-ray sample size: A robust approach for different medical imbalanced scenarios. Applied Sciences. 2022;12(8):3903.
32. Ning W, Lei S, Yang J, Cao Y, Jiang P, Yang Q, et al. Open resource of clinical data from patients with pneumonia for the prediction of COVID-19 outcomes via deep learning. Nature biomedical engineering. 2020;4(12):1197-207.